\documentclass[12pt,letterpaper]{article}
\usepackage[top=0.85in,left=0.85in,footskip=0.75in,marginparwidth=1in]{geometry}
\usepackage{amsmath}
\usepackage{upgreek}
\usepackage{xcolor}
\usepackage{graphicx}
\usepackage{upgreek}

\usepackage[utf8]{inputenc}

\usepackage{cite}

\usepackage{nameref,hyperref}

\usepackage{microtype}
\DisableLigatures[f]{encoding = *, family = * }

\raggedright
\usepackage{changepage}

\usepackage[aboveskip=1pt,labelfont=bf,labelsep=period,singlelinecheck=off]{caption}

\makeatletter
\renewcommand{\@biblabel}[1]{\quad#1.}
\makeatother

\usepackage{lastpage,fancyhdr,graphicx}
\usepackage{epstopdf}
\fancyheadoffset[L]{2.25in}
\fancyfootoffset[L]{2.25in}

\usepackage{color}

\definecolor{Gray}{gray}{.25}

\usepackage{graphicx}

\usepackage{sidecap}

\usepackage{wrapfig,cite}
\usepackage[pscoord]{eso-pic}
\usepackage[fulladjust]{marginnote}
\reversemarginpar

\usepackage{ragged2e}
\justifying\let\raggedright\justifying

\begin{document}

\vspace*{0.35in}

\begin{flushleft}
{\Large
\textbf\newline{Slow wave and truly rainbow trapping in one-way terahertz waveguide}
}
\newline

Jie Xu,\textsuperscript{1} Panpan He,\textsuperscript{2} Delong Feng,\textsuperscript{1} Kangle Yong,\textsuperscript{1} Lujun Hong,\textsuperscript{3} Yun Shen,\textsuperscript{3} and Yun Zhou,\textsuperscript{1,*}

\bigskip
\bf{1} School of Medical Information and Engineering, Southwest Medical University, Luzhou 646000, China\\
\bf{2} Department of Electronic Engineering, Luzhou Vocational and Technical College, Luzhou 646000, China\\
\bf{3} Institute of Space Science and Technology, Nanchang University, Nanchang 330031, China\\

\bigskip
* 18591988761@163.com


\end{flushleft}

\section*{Abstract}
Slow or even stop electromagnetic (EM) waves attract researchers' attentions for its potential applications in energy storage, optical buffer and nonlinearity enhancement. However, in most cases of the EM waves trapping, the EM waves are not truly trapped due to the existence of reflection. In this paper, a novel metal-semiconductor-semiconductor-metal (MSSM) structure, and a novel truly rainbow trapping in a tapered MSSM model at terahertz frequencies are demonstrated by theoretical analysis and numerical simulations. More importantly, functional devices such as optical buffer, optical switch and optical filter are achieved in our designed MSSM structure based on truly rainbow trapping theory. Owing to the property of one-way propagation, these new types of optical devices can be high-performance and are expected to be used in integrated optical circuits.


\section{Introduction}
Electromagnetic (EM) waves can propagate in both forward and backward directions in a general physical system because of the presence of the time-reversal symmetry of the system. One-way propagating EM modes, on the other hand, are those EM waves that can only travel in one direction in a physical system, which is similar to the chiral edge state in the quantum Hall effect\cite{Prang:Th}. The unidirectional EM waves can bypass imperfections as well as bends because they are immune to backscattering. Applying an external magnetic field is one of the fundamental ways to break the time-reversal symmetry and further designing one-way waveguides\cite{Wang:Re,Ao:On,Yu:On,Hu:Br}. It was reported that the one-way EM waves are found when an external magnetic field is applied to two-dimensional photonic crystals composed of magneto-optical (MO) materials due to broken time-reversal symmetry\cite{Raghu:An}. In 2009, the existence of such one-way EM modes was experimentally observed in a photonic crystal consisted of yttriun-iron-garnet (YIG) which is a MO material in microwave regime\cite{Wang:Ob}. One-way EM modes sustained by the interface of dielectric and MO material is called the surface magnetoplasmon (SMP)\cite{Brion:Th,Wallis:Th}, and because of the dc magnetic field, the asymptotic frequencies of SMPs are different in forward and backward. Thus, there are a complete one-way propagating (COWP) band for SMPs\cite{Shen:Ba}. Recently, we presented Y-branch\cite{Zou:Hi} and t-shaped\cite{Hong:Hi} beam splitters based on one-way SMPs. More recently, time-bandwidth limit was reported to be broken in a truncated one-way waveguide in which SMPs are localized and behave as in a 'zero-dimensional' cavity\cite{Tsakmakidis:Br} around the terminal.

Slowing wave or light has been a research hotspot for the past two decades. In 1999, an experimental demonstration of slowing the speed of light to 17 metres per second was present by utilizing a quantum effect called electromagnetically induced transparency (EIT)\cite{Hau:Li}. In addition to EIT, it was also possible to use photonic crystals (PhCs) or metamaterials to slow light. By tailoring the dispersion properties of light in the photonic bandgaps (PBGs) of PhCs, the group velocity of light can be dramatically decreased \cite{Schulz:Di,Ek:Sl,Schulz:Ph,Munoz:Op,Yan:Sl}. In the other hand, metamaterials can possess special properties like negative refractive index, which can also be used in slowing EM waves or light\cite{Tsakmakidis:Tr}. Graphene, undoubtedly, is in the spotlight today, and graphene-based metamaterials can slow light using the principle of plasmon-induced transparency (PIT) which is similar to EIT\cite{Zhang:Pl,Zhang:Ab,Gao:Du}. Rainbow trapping refers to the capture of EM waves or light of different frequencies at different spatial locations. There are various structures in which the rainbow trapping can be achieved, including three-dimensional photonic crystals with spatially distance between layers changing along the directions of propagation\cite{Hayran:Ra}, chirped plasmonic waveguide\cite{Chen:Ra}, metasurfaces\cite{Xuz:Ra}, metamaterials\cite{Tsakmakidis:Tr,Hu:Ra}, and one-way waveguides consisting of MO materials\cite{Liu:Tr,Xu:Br}.

Rainbow trapping is promising in optical storage, however, due to the coupling between forward and backward EM modes, the slowed down or trapped fields could gradually propagate to opposite direction\cite{He:Re}. For truly rainbow trapping, Ref. \cite{Liu:Tr} presented a metal-dielectric-YIG structure under an external magnetic field linearly increasing along the direction of the EM propagation in the microwave regime. In this paper, we achieve the rainbow trapping in terahertz regime based on a novel mechanism in a metal-semiconductor-semiconductor-metal (MSSM) model. Instead of using linearly changing dc magnetic field, we achieve the rainbow trapping in a tapered MSSM structure, which makes our model more practical than the one in Ref. \cite{Liu:Tr}. Besides, we further present potential ways to utilize our MSSM model in optical buffer, filter and switch.

\section{Three complete one-way propagation bands}

\begin{figure}[ht]
	\centering\includegraphics[width=4.5 in]{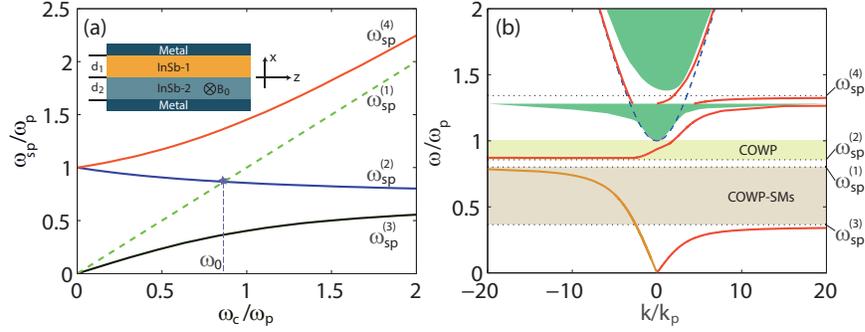}
	\caption{ (a) Four asymptotic frequencies (AFs) of SMPs as functions of $\omega_{\rm c}$. The inset illustrates the schematic of the metal-semiconductor-semiconductor-metal (MSSM) model. (b) The dispersion diagram of SMPs (red solid lines) and SMs (yellow solid line) as $d_1=d_2=0.1 \lambda_{\rm p}$ and $\omega_{\rm c}=0.8\omega_{\rm p}$. Four horizontal lines represent the AFs and the green shaded areas are the bulk zones in InSb-2.}\label{Fig1}
\end{figure}

The inset of Fig. 1(a) shows the physical model of the metal-semiconductor-semiconductor-metal (MSSM) structure, in which one of semiconductors is assumed to be under an external magnetic field $B_0$ with -y direction. Because of the existence of $B_0$, the time-reversal symmetry in the MSSM waveguide will be broken. We first theoretically study the dispersion and propagation properties of the electromagnetic (EM) waves in the MSSM model. The metal can be treated as perfect electric conductor (PEC) in the terahertz regime\cite{Shen:On} and in this paper, the semiconductors are assumed to be indium antimonide (InSb). Moreover, to distinguish two layers of InSb, they are named InSb-1 and InSb-2. The upper layer semiconductor (InSb-1) is described by the Drude model, thus its relative permittivity can be written as

\begin{equation}
	\varepsilon_{\rm Drude}=\varepsilon_\infty\left(1-\frac{\omega_{\rm p}^2}{\omega^2}\right),
\end{equation}
where $\varepsilon_\infty$, $\omega_{\rm p}$ and $\omega$ are the high-frequency (relative) permittivity, the plasma frequency and the angular frequency, respectively. The relative permittivity of Insb-2, as reported in our earlier paper, has off-diagonal tensor form

\begin{eqnarray}
	\mathop \varepsilon \limits^ \leftrightarrow  = \left[ {\begin{array}{*{20}{c}}
			{{\varepsilon _1}}&0&{i{\varepsilon _2}}\\
			0&{{\varepsilon _3}}&0\\
			{ - i{\varepsilon _2}}&0&{{\varepsilon _1}}
	\end{array}} \right],
\end{eqnarray}
with
$\varepsilon_{1}  = \varepsilon _\infty  \left( {1 - \frac{{\omega
			_{\rm p}^2 }}{{\omega ^2  - \omega _{\rm c}^2 }}} \right),
\varepsilon_{2}  = \varepsilon _\infty  \frac{{\omega_{\rm c}\omega _{\rm p}^2
}}{{\omega (\omega ^2  - \omega _{\rm c}^2 )}},
\varepsilon_{3}  = \varepsilon _\infty  \left( {1 - \frac{{\omega
			_{\rm p}^2 }}{{\omega ^2 }}} \right)$,
where $\omega_{\rm c}=eB_0/m^*$ is the electron cyclotron frequency. $e$ and $m^*$ are the charge and effective mass of an electron, respectively. Combining Maxwell's equations and all the boundary conditions, one can easily obtain the dispersion equation of surface waves in this MSSM waveguide as written below

\begin{eqnarray}
	\left( {{k^2} - {\varepsilon _1}k_0^2} \right)\tanh ({\alpha _{\rm s}}{d_2}) + \frac{{{\varepsilon _1}}}{{{\varepsilon _{\rm Drude}}}}{\alpha _D}\left[ {{\alpha _{\rm s}} - \frac{{{\varepsilon _2}}}{{{\varepsilon _1}}}k\tanh ({\alpha _{\rm s}}{d_2})} \right]\tanh ({\alpha _D}{d_1}) = 0,
\end{eqnarray}
where $d_1$ ($d_2$) is the thickness of InSb-1 (InSb-2). $k$, $k_0$, $\alpha_D=\sqrt{k^2-\varepsilon_{\rm Drude}k_0^2}$ and $\alpha_{\rm s}=\sqrt{k^2-\varepsilon_{\rm v}k_0^2}$ ($\varepsilon_{\rm v}=\varepsilon_1-\varepsilon_2^2/\varepsilon_1$ is the Voigt permittivity) are the propagation constant, the wavenumber in vacuum, the attenuation coefficient of surface EM waves in InSb-1 and InSb-2, respectively. From Eq. (3) we find $\varepsilon_1+\varepsilon_{\rm Drude}-\varepsilon_2=0$ for $k\rightarrow +\infty$. Thus, we have
\begin{eqnarray}
	2\bar{\omega}^3-2\bar{\omega_{\rm c}}\bar{\omega}^2-2\bar{\omega}+\bar{\omega_{\rm c}}=0,
\end{eqnarray}
where $\bar{\omega}=\omega/\omega_{\rm p}$ and $\bar{\omega_{\rm c}}=\omega_{\rm c}/\omega_{\rm p}$. To analyse the roots of Eq. (4), we set $a=2$, $b=-2\bar{\omega_{\rm c}}$, $c=2$ and $d=\bar{\omega_{\rm c}}$. By using Shengjin's Formulas, we have $A=b^2-3ac=4\bar{\omega_{\rm c}}^2+12$, $B=bc-9ad=-14\bar{\omega_{\rm c}}$ and $C=c^2-3bd=4+6\bar{\omega_{\rm c}}^2$. As a result, the discriminant of Shengjin's Formulas $\Delta=B^2-4AC$ remains negative for all values of $\bar{\omega_{\rm c}}$ and there must be three different solutions for Eq. (4). Further more, the product of three roots equals $d$ and $d=\bar{\omega_{\rm c}}>0$, which implys one of the roots must be negative. Therefore, for $k>0$, there are two asymptotic frequencies (AFs) for SMPs. Likewise, as $k\rightarrow -\infty$, Eq. 3 changes to
\begin{eqnarray}
	(2\bar{\omega}^3+2\bar{\omega_{\rm c}}\bar{\omega}^2-2\bar{\omega}-\bar{\omega_{\rm c}})(\bar{\omega}-\bar{\omega_{\rm c}})=0,
\end{eqnarray}
$\bar{\omega}=\bar{\omega_{\rm c}}$ is one of the root of Eq. 5 and for the second form, $\Delta<0$ and $d=-\bar{\omega_{\rm c}}<0$, which indicates that there should be two or four AFs in the MSSM model for $k<0$. By using numerical calculations, we plot the AFs in Fig. 1(a) as a function of $\omega_{\rm c}$, and we found that there exist four different AFs in total when $0<\bar{\omega_{\rm c}}<2$. The AFs are named $\omega_{\rm sp}^{(1)}$, $\omega_{\rm sp}^{(2)}$, $\omega_{\rm sp}^{(3)}$ and $\omega_{\rm sp}^{(4)}$. Such special dispersion property can also be seen in Fig. 1(b), in which $d_1=d_2=0.1\lambda_{\rm p}$ and $\omega_{\rm c}=0.8\omega_{\rm p}$. The parameters of InSb are $\varepsilon_\infty=15.6$ and $\omega_{\rm p}=4\pi\times10^{12}$ $\rm rad/m$\cite{Isaac:De}. In Fig. 1(b), the green coloured zones and the blue dashed curve represent the bulk zones of InSb-2 and InSb-1, respectively. The yellow solid line represent a special kind of SMPs, which is sustained by semiconductor-metal interface, we named it SMs to distinguish from regular SMPs. $\omega_{\rm sp}^{(1)}=\omega_{\rm c}$ is the AF of SMs, and the other three horizontal dotted lines in Fig. 1(b) are the AFs of SMPs. The yellow and gray coloured zones are complete one-way propagation (COWP) bands. Note that, when $\omega_{\rm c}>\omega_0$ ($\omega_0\approx0.863\omega_{\rm p}$) interesting phenomena will appear. However, in this section, we only consider the conditions of $\omega_{\rm c}<\omega_0$ for studying one-way propagation properties in the MSSM structure. 

\begin{figure}[pt]
	\centering\includegraphics[width=4.5 in]{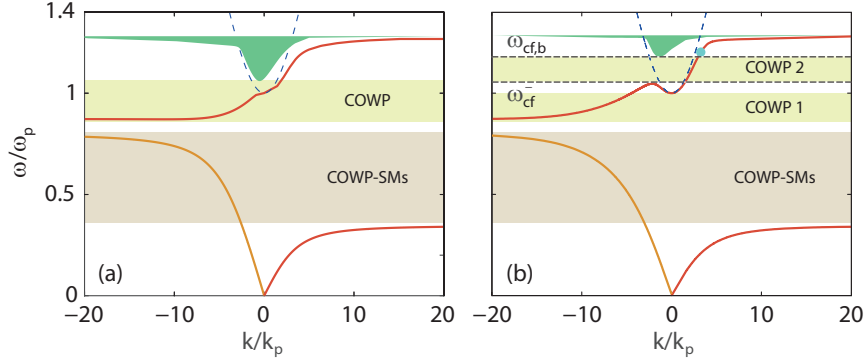}
	\caption{ The dispersion diagram of the MSSM configuration for the cases of (a) $d_2=0.06\lambda_{\rm p}$ and (b) $d_2=0.03\lambda_{\rm p}$. Two shaded yellow areas and the horizontal dashed lines respectively represent the COWP bands of SMPs and the cut-off frequencies of the bulk zone and the dispersion curves as $k\leq 0$. The grey coloured area represents the COWP band of SMs. The other parameters are $d_1=0.1\lambda_{\rm p}$, $\omega_{\rm c}=0.8\omega_{\rm p}$ and $\varepsilon_\infty=15.6$.}\label{Fig2}
\end{figure}

\begin{figure}[pt]
	\centering\includegraphics[width=4.5 in]{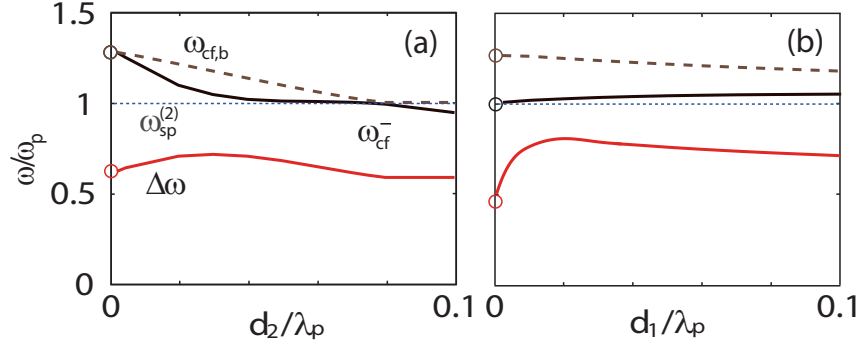}
	\caption{ The total bandwidth $\Delta \omega$ of the COWP bands as functions of (a) $d_2$ with $d_1=0.1\lambda_{\rm p}$ and (b) $d_1$ with $d_2=0.03\lambda_{\rm p}$. The other parameters are the same as in Fig. 2.}\label{Fig3}
\end{figure}

\begin{figure}[ht]
	\centering\includegraphics[width=4 in]{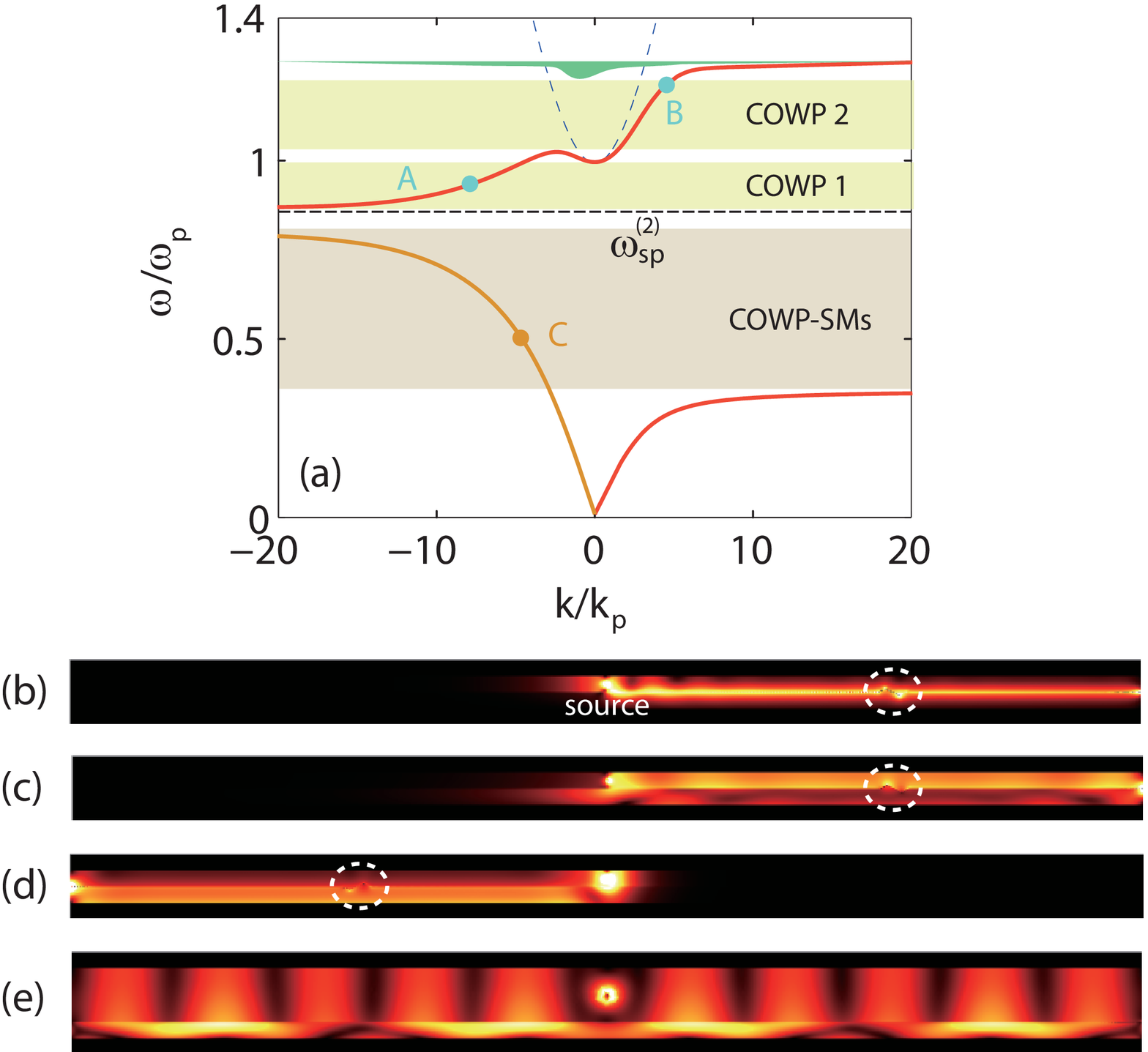}
	\caption{ (a) The dispersion diagram of SMPs and SMs. (b)-(d) The simulated electric field distributions in the MSSM waveguide with 'zig-zag' interfaces emphasized by the white circle for three working frequencies, i.e. (b) $f=0.9f_{\rm p}$ (marked by point A in (a)), (c) $f=1.2f_{\rm p}$ (marked by point B in (a)) and (d) $f=0.5f_{\rm p}$ (marked by point C in (a)). The thickness parameters in (a)-(d) are $d_1=d_2=0.03\lambda_{\rm p}$. (e) The simulated electric field distribution as $d_2=0.1\lambda_{\rm p}$, $d_1=0.03\lambda_{\rm p}$, and $f=1.2f_{\rm p}$ (marked by point in Fig. 2(b)). The other parameters are $\omega_{\rm c}=0.8\omega_{\rm p}$ and the electron scattering frequency of lossy InSb is $\nu=0.001\omega_{\rm p}$.}\label{Fig4}
\end{figure}

In our previous work, we found that by cutting off InSb-2 in the vertical direction could significantly enlarge the bandwidth of the COWP band of SMPs\cite{Xu:Br}. Thus, we plot the dispersion diagram of SMPs for two different values of $d_2$ in Fig. 2. The thicknesses of InSb-2 are respectively $d_2=0.06\lambda_{\rm p}$ (Fig. 2(a)) and $d_2=0.03\lambda_{\rm p}$ (Fig. 2(b)). The dispersion curves of SMPs in Fig. 2(a) are quite similar to the one in Fig. 1(b), except for the region around $\omega=\omega_{\rm p}$. The total bandwidth of the COWP bands ($\Delta \omega$) in Fig. 2(a) is obviously large than the one in Fig. 1(b) since the edge of the bulk zone (the green coloured areas) of InSb-2 raised up when $d_2$ decreased. Note that the COWP-SMs remains the same in Fig. 1(b) and Fig. 2. More interestingly, when $d_2$ is small enough (e.g. $d_2=0.03\lambda_{\rm p}$), the COWP band of SMPs is divided up into two COWP bands (see Fig. 2(b)), i.e. COWP-1 and COWP-2. The COWP-2 is limited by the cut-off frequency ($\omega_{\rm cf,b}$) of bulk zones of InSb-2 and the cutoff frequency ($\omega_{\rm cf}^-$) of SMPs for $k\leq0$. We can not tell from the equations whether $\Delta \omega$ keeps getting bigger as $d_2$ decrease. To make clear the relation between $d_1$, $d_2$ and $\Delta\omega$, we did numerical calculation as $\omega_{\rm c}=0.8\omega_{\rm p}$ and the results are presented in Fig. 3. In Fig. 3(a), $d_1=0.1\lambda_{\rm p}$ and $d_2$ is changed from 0 to $0.1\lambda_{\rm p}$. For $d_1=d_2=0.1\lambda_{\rm p}$, $\omega_{\rm cf,b}=\omega_{\rm p}$ and $\omega_{\rm cf}^-<\omega_{\rm sp}^{(2)}$, which fit well with the results shown Fig. 1(b). Moreover, $\Delta\omega$ (red line) will not increase all the time when $d_2$ decrease, and it reaches a maximum $\Delta\omega_{\rm max}\approx0.71\omega_{\rm p}$ around $d_2= 0.03\lambda_{\rm p}$. Furthermore, the relation between $\Delta \omega$ and $d_1$ are also studied for $d_2=0.03\lambda_{\rm p}$. As shown in Fig. 3(b), due to the thin InSb-2, $\omega_{\rm cf}^-$ is constantly larger than $\omega_{\rm sp}^{(2)}=\omega_{\rm p}$, and $\Delta\omega_{\rm max}\approx0.79\omega_{\rm p}$ around $d_1=0.03\lambda_{\rm p}$. We note that the decrease of $\Delta \omega$ as $d_1$ continue decreasing from $0.03\lambda_{\rm p}$ is due to the raising up peak of the branch of the dispersion curves of SMPs (e.g. the red line at bottom right of Fig. 2). When designing subwavelength devices with $d_1<0.03\lambda_{\rm p}$, this phenomenon should be considered. However, we just study the structures of $d_1\geq0.03\lambda_{\rm p}$ in this paper.

Fig. 4(a) shows the three COWP bands of the surface EM modes for $d_1=d_2=0.03\lambda_{\rm p}$ and $\omega_{\rm c}=0.8\omega_{\rm p}$, and three working frequencies (marked by points A, B and C) are chosen to perform numerical simulations by using finite element method (FEM). We emphasise here, in this paper, the lossy InSb with the scattering frequency $\nu=0.001\omega_{\rm p}$ are considered in numerical simulations. Figs. 4(b)-(d) are the simulated electric field distributions as $f=0.935f_{\rm p}$ ($f_{\rm p}=2$ THz), $f=1.2f_{\rm p}$ and $f=0.5f_{\rm p}$, respectively. The terminals of the MSSM waveguide are assumed to be PEC wall. As a result, the excited EM waves in Figs. 4(b)-(d) are allowed to propagate in only one direction and no reflection was found in these simulations, which means all three frequencies are in the COWP band. Moreover, for thicker InSb-2 case, e.g. $d_1=0.1\lambda_{\rm p}$, the working frequency $f=1.2f_{\rm p}$ will be no longer in the COWP band (see Fig. 2(b)). Fig. 4(e) shows the excited EM mode with $f=1.2f_{\rm p}$ propagating to both forward and backward direction. We note here that it is the first time the one-way SMs and the one-way (regular) SMPs are simultaneously found in a physical system, and we believe that due to the different characteristics between two unidirectional EM modes, they can be used in different ways, for example, the one-way SMPs can be used to combine the one-way waveguide and the normal dielectric waveguide\cite{Shen:On}, and the one-way SMs are promising for backscattering immune EM waves propagation and isolation when considering nonlocal effects\cite{Gangaraj:Do}.

\section{Two types of rainbow trapping}

\begin{figure}[ht]
	\centering\includegraphics[width=4.5 in]{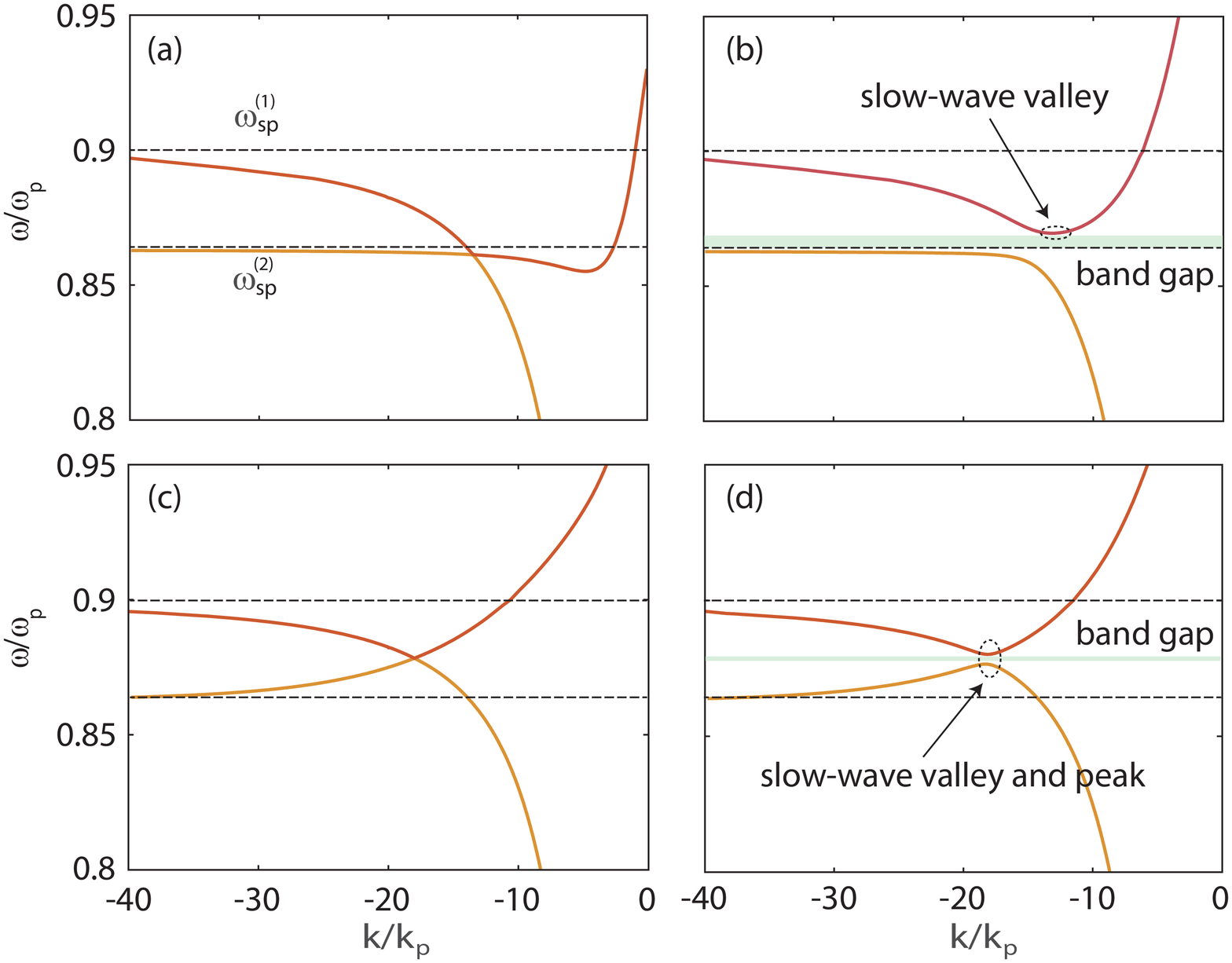}
	\caption{ Zoomed in dispersion diagram of SMPs and SMs when $\omega_{\rm c}=0.9\omega_{\rm p}$ and the thickness parameters are (a) $d_1=d_2=0.1\lambda_{\rm p}$, (b)$d_1=0.1\lambda_{\rm p}$, $d_2=0.05\lambda_{\rm p}$, (c) $d_1=0.01\lambda_{\rm p}$, $d_2=0.1\lambda_{\rm p}$, and (d) $d_1=0.01\lambda_{\rm p}$, $d_2=0.05\lambda_{\rm p}$. The horizontal lines represent two asymptotic frequencies $\omega_{\rm sp}^{(1)}$ and $\omega_{\rm sp}^{(2)}$. The shaded areas in (b) and (d) are the band gaps of the MSSM waveguide.}\label{Fig5}
\end{figure}

Fig. 1 tells that when the external magnetic field $B_0$ is small ($\omega_{\rm c}<\omega_0$), two kinds of the surface EM modes, i.e. regular SMPs and SMs will not couple with each other for there is a band gap limited by two asymptotic frequencies $\omega_{\rm sp}^{(1)}$ and $\omega_{\rm sp}^{(2)}$. However, once $B_0$ is large enough ($\omega_{\rm c}>\omega_0$), the band gap should disappear and new phenomenons may appear. For this reason, we studied the dispersion diagrams around the 'gap' as $\omega_{\rm c}=0.9\omega_{\rm p}$. The AFs, in this condition, are respectively $\omega_{\rm sp}^{(1)}=0.9\omega_{\rm p}$ and $\omega_{\rm sp}^{(2)}=0.863\omega_{\rm p}$. Fig. 5 shows the dispersion diagram for four different cases and the according thickness parameters are (a) $d_1=d_2=0.1\lambda_{\rm p}$, (b)$d_1=0.1\lambda_{\rm p}$, $d_2=0.05\lambda_{\rm p}$, (c) $d_1=0.01\lambda_{\rm p}$, $d_2=0.1\lambda_{\rm p}$, and (d) $d_1=0.01\lambda_{\rm p}$, $d_2=0.05\lambda_{\rm p}$. The strong couple effects of SMPs and SMs can be seen in Fig. 5(a) and (c), in which $d_2=0.1\lambda_{\rm p}$ and the dispersion curves of SMs (the yellow line) and SMPs (the red line) cross each other. More interestingly, slow-wave valley or peak is found in Figs. 5(b) and 5(d), as well as new band gaps. It is obvious that when the thickness of InSb-2 ($d_2$) decreas from $0.1\lambda_{\rm p}$ to $0.05\lambda_{\rm p}$, the dispersion curves of SMs are nearly unchanged and the dispersion curves of SMPs shift up. The unchanged and shift up branches of dispersion curves cause the slow-wave valley (peak) and the band gaps in Figs. 5(b) and 5(d). In addition, as we discussed in Section 2, $d_1=0.01\lambda_{\rm p}$ is not suitable parameter for design and application, we assume $d_1=0.1\lambda_{\rm p}$ for slow-wave and rainbow trapping study.

\begin{figure}[ht]
	\centering\includegraphics[width=4 in]{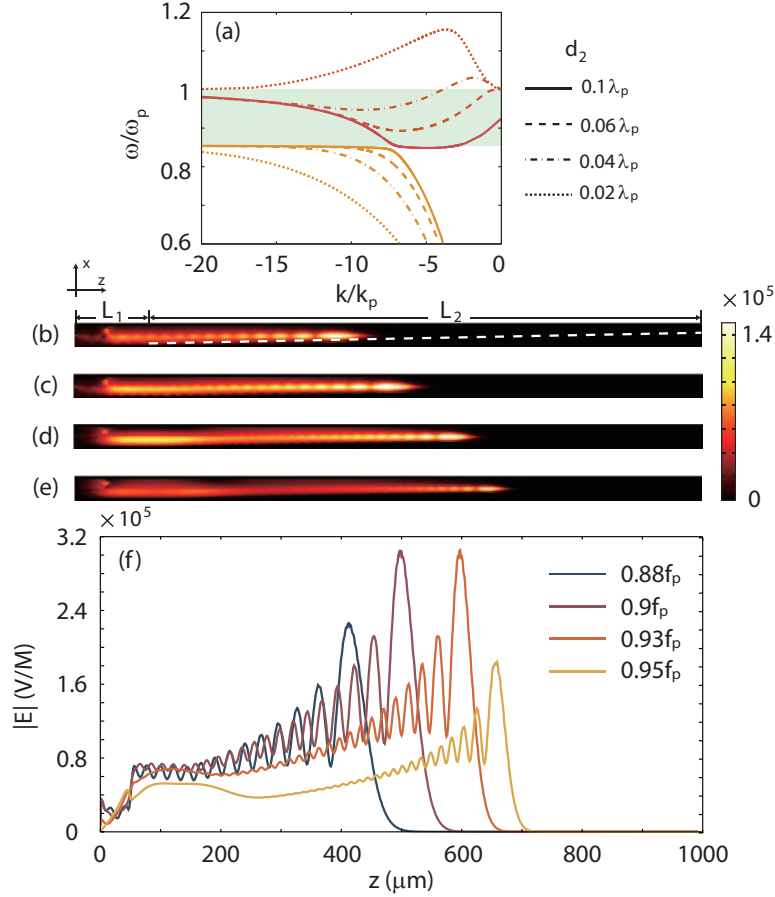}
	\caption{ (a) The dispersion curves of surface waves for four different value of $d_2$ as $d_1=0.1\lambda_{\rm p}$ and $\omega_{\rm c}=\omega_{\rm p}$. (b)-(e) Simulated electric field amplitudes in a tapered MSSM configuration for (b) $f=0.88f_{\rm p}$, (c) $f=0.9f_{\rm p}$, (d) $f=0.93f_{\rm p}$ and (e) $f=0.95f_{\rm p}$. The lengthes of the straight and tapered parts of the MSSM waveguide are respectively $L_1=150\upmu$m and $L_2=850\upmu$m, and the thickness parametes of the straight part are $d_1=d_2=0.1\lambda_{\rm p}$. (f) The electric field amplitude distributions along the InSb-InSb interfaces in (b), (c), (d) and (e).}\label{Fig6}
\end{figure}

\begin{figure}[ht]
	\centering\includegraphics[width=4 in]{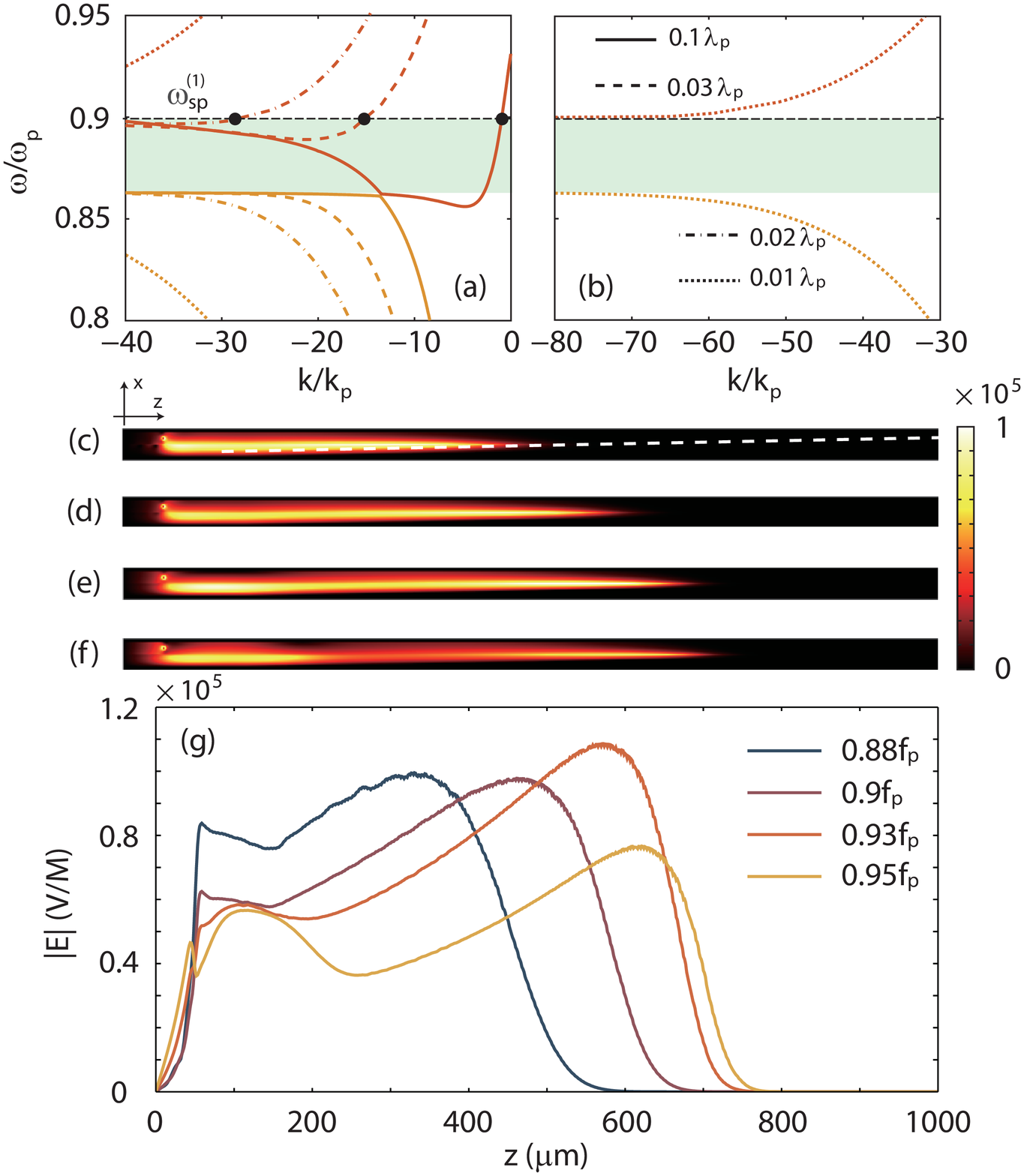}
	\caption{ Truly rainbow trapping. (a), (b) The dispersion diagrams for four different $d_2$ (as shown the inset of (b)). The other parameters are $d_1=0.1\lambda_{\rm p}$ and $\omega_{\rm c}=0.9\omega_{\rm p}$. (c)-(f) The simulated electric field distribution in the same tapered structure as in Fig. 6(b). $f/f_{\rm p}=\omega_{\rm c}/\omega_{\rm p}$ and the working frequencies in (c)-(f) are the same as in Figs. 6(b)-(e). (g) The electric field amplitudes distribution along the InSb-InSb interfaces in (c), (d), (e) and (f).}\label{Fig7}
\end{figure}

The thickness-based slow-wave valley can be used to achieve rainbow trapping. Fig. 6(a) shows the dispersion diagrams for four different values of $d_2$ and the solid lines, dashed lines, chain-dotted lines and dotted lines represent the dispersion relation in the MSSM structures with $d_2=0.1\lambda_{\rm p}$, $d_2=0.06\lambda_{\rm p}$, $d_2=0.04\lambda_{\rm p}$ and $d_2=0.02\lambda_{\rm p}$, respectively. It is clearly shown in Fig. 6(a) that, when $d_2$ decrease from $0.1\lambda_{\rm p}$ (solid lines) to $0.02\lambda_{\rm p}$ (dotted lines), the cut-off frequency of SMs (yellow lines) do not change and the slow-wave valley gradually changed to slow-wave peak (red dotted line) with the cut-off frequency shifts up to $\omega_{\rm sp}>\omega_{\rm c}$ ($\omega_{\rm c}=\omega_{\rm p}$). We further designed a straight-tapered joint MSSM waveguide to trap rainbow. As shown in Fig. 6(b), the lengthes of the straight and tapered parts of the MSSM waveguide are set to be $L_1=150\upmu$m and $L_2=850 \upmu$m, respectively. In the tapered part, we set a metal layer (the white dashed line in Fig. 6(b)) in the InSb-2 layer to linearly decrease $d_2$. Figs. 6(b)-(e) are the electric field distribution of the simulations and, correspondingly, the working frequencies are $f=0.88f_{\rm p}$, $f=0.9f_{\rm p}$, $f=0.93f_{\rm p}$ and $f=0.95f_{\rm p}$. One can clearly see the trapped rainbow in Figs. 6(b)-(d) since the EM waves with higher the frequency the longer the trapping length. More visual results are shown in Fig. 6(f), in which the four lines represent the amplitudes of the electric field along the InSb-InSb interface. Unlike the electric field distribution in one-way propagation cases, the electric fields in Figs. 6(b)-(d) are not uniform and there obviously exist reflections and interference of EM waves in the waveguide. According to the simulation results, this slow-wave valley-induced rainbow trapping has much higher trapping efficiency than the slow-wave peak-induced rainbow trapping reported in our previous work\cite{Xu:Br} for there are no SMPs for $k>0$ around slow-wave valley. Besides, such straight-tapered joint MSSM configuration can also be used to achieve truly rainbow trapping. 

The concept of truly trapped rainbow comes from Ref. \cite{He:Re}, in which the researcher illustrates the coupling between forward and backward propagating EM modes in a tapered metamaterial waveguide will greatly influence the process of rainbow capture. The critical problem in truly rainbow trapping is how to overcome the forward-backward coupling. Here, we propose a novel truly rainbow trapping theory based on the MSSM configuration. Fig. 5 and Fig. 6(a) show that once $\omega_{\rm c}>\omega_0$ the disperison curves of SMPs (red lines) will shift up with the decrease of $d_2$. When $d_2$ is small enough, the dispersion curves of SMPs can 'escape' to $\omega>\omega_{\rm sp}^{(1)}$ region (e.g. red dotted line in Fig. 6(a)). Much more interestingly, $\omega=\omega_{\rm sp}^{(1)}$ appears to be a solution of Eq. (3) in all cases. Figs. 7(a) and 7(b) provide a more clear description of truly rainbow trapping theory, in which the parameters are $d_1=0.1\lambda_{\rm p}$ and $\omega_{\rm sp}^{(1)}=\omega_{\rm c}=0.9\omega_{\rm p}$. Four types of lines represent four values of $d_2$ and they are respectively $d_2=0.1\lambda_{\rm p}$ (solid line), $d_2=0.03\lambda_{\rm p}$ (dash line) and $d_2=0.02\lambda_{\rm p}$ (chain-dotted line). It is quit clear that $f/f_{\rm p}=\omega_{\rm sp}^{(1)}/\omega_{\rm p}$ (black points in Fig. 7(a)) holds in three conditions, which means the EM waves with $f=\omega_{\rm sp}^{(1)}/\omega_{\rm p} \cdot f_{\rm p}$ can propagate in the corresponding structures and are prohibited in the case of $d_2=0.01\lambda_{\rm p}$. Based on this theory, we perfume full wave simulations in the same waveguide as the one in Fig. 6(b). In Figs. 7(c)-(f), the working frequencies f and the electron cyclotron frequency $\omega_{\rm c}$ satisfied the equation $f/f_{\rm p}=\omega_{\rm c}/\omega_{\rm p}$. Four working frequencies are the same with those in Figs. 6(b)-(e). As shown in Figs. 7(c)-(f), the EM waves with different frequencies are trapped at different locations and, excitingly, the electric fields in all simulations have uniform distributions. Fig. 7(g) indicates the according amplitude of the electric field along the InSb-InSb interface in the simulations and there is nearly no reflection and interference. Comparing with the trapped rainbow mentioned in Fig. 6, the rainbow in Fig. 7 is truly trapped for the trapped EM waves can not travel in the reverse direction.


\section{Optical filter, switch, and buffer based on the MSSM configuration}
\begin{figure}[ht]
	\centering\includegraphics[width=4 in]{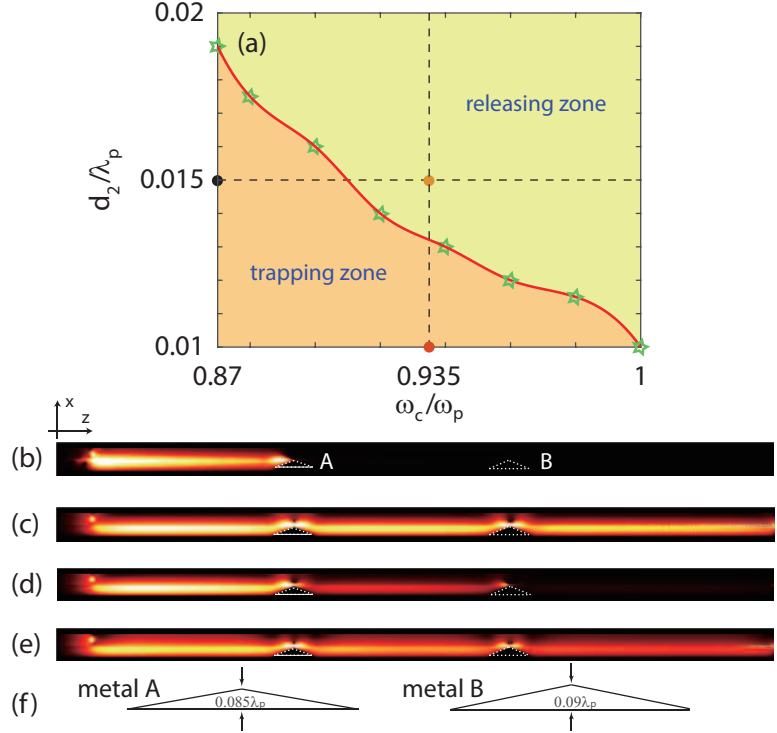}
	\caption{ (a) The mechanism of trapping and releasing the EM wave based on $f/f_{\rm p}=\omega_{\rm c}/\omega_{\rm p}$. (b)-(e) The numerical simulations in a MSSM structure with two metal barriers A and B (the white lines) for (b) $f=0.87f_{\rm p}$ and $\omega_{\rm c}=0.87\omega_{\rm p}$, (c) $f=0.935f_{\rm p}$ and $\omega_{\rm c}=0.87\omega_{\rm p}$, (d) $f=0.935f_{\rm p}$ and $\omega_{\rm c}=0.935\omega_{\rm p}$ and (e) $f=0.95f_{\rm p}$ and $\omega_{\rm c}=0.935\omega_{\rm p}$. (f) The schematic of metal barriers. In the straight part of the waveguide, the thickness parameters are $d_1=d_2=0.1\lambda_{\rm p}$.}\label{Fig8}
\end{figure}
\begin{figure}[ht]
	\centering\includegraphics[width=5 in]{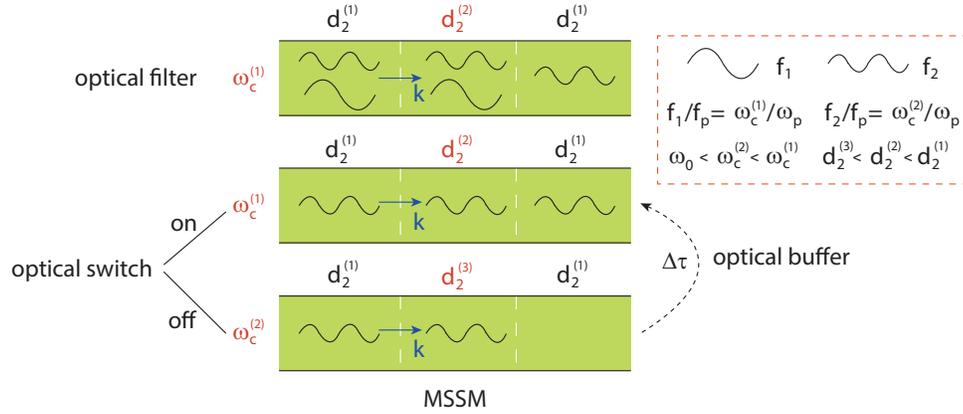}
	\caption{ Realization of optical filter, optical switch and optical buffer based on the MSSM configuration.}\label{Fig9}
\end{figure}
Truly rainbow trapping has many potential applications such as energy storage, optical signal processing and optical nonlinearity enhancement. In this section, we design an optical device which can be used as an optical buffer, optical filter and optical switch as well. We first study the relation between $d_2$ and $\omega_{\rm c}$ as $d_1=0.1\lambda_{\rm p}$ under the condition of $f/f_{\rm p}=\omega_{\rm c}/\omega_{\rm p}$. For truly trapping rainbow, the external magnetic field must be relatively large, more specifically, $\omega_{\rm c}>\omega_0$. Thus, in this section, we only consider $0.87\omega_{\rm p}\leq\omega_{\rm c}\leq\omega_{\rm p}$. In Fig. 8(a), the $d_2-\omega_{\rm c}$ space is splitted into two areas by the red line which is the fitted curve of numerical solutions (marked by stars) of Eq. 3 when truly rainbow trapping happens. The yellow coloured area and the orange coloured area are named the releasing zone and the trapping zone. Points located in the trapping or releasing zone implys the EM waves with $f=\omega_{\rm c}/\omega_{\rm p} \cdot f_{\rm p}$ can or can not be trapped in the corresponding condition. Moreover, the y axis of Fig. 8(a), the vertical dashed line, the horizontal dashed line and the x axis of Fig. 8(a) represent $\omega_{\rm c}=0.87\omega_{\rm p}$, $\omega_{\rm c}=0.935\omega_{\rm p}$, $d_2=0.015\lambda_{\rm p}$ and $d_2=0.01\lambda_{\rm p}$. According to the zones of trapping and releasing, the black or the red point implys that, when $f/f_{\rm p}=\omega_{\rm c}/\omega_{\rm p}$, the EM waves can be trapped at the location with $d_2=0.015\lambda_{\rm p}$ or $d_2=0.01\lambda_{\rm p}$. On the other hand, the orange point means the EM waves cannot be trapped in such case. Thus, by carefully designing the thickness of the InSb-2 layer and by controlling the external magnetic field, we can trap or release the EM waves with specific frequencies in a MSSM model. Figs. 8(b)-(d) shows the full wave simulations results in a MSSM waveguide with two triangular metal barriers (white lines) A and B which are put on the bottom metal layer of the MSSM waveguide. The positions of the source, peaks of metals A and B are $z=50\upmu$m, $z=330\upmu$m and $z=630\upmu$m, respectively. $d_1=d_2=0.1\lambda_{\rm p}$ in the straight part of the MSSM waveguide. Moreover, the height of two metal barriers are $0.085\lambda_{\rm p}$ and $0.09\lambda_{\rm p}$, respectively, corresponding to $d_2=0.015\lambda_{\rm p}$ (the horizontal line in Fig. 8(a)) and $d_2=0.01\lambda_{\rm p}$ (the x axis of Fig. 8(a)). In Fig. 8(b), we set $f/f_{\rm p}=\omega_{\rm c}/\omega_{\rm p}=0.87$. As a result, the excited EM wave is trapped before it encounters the peak of metal A. Comparing with Fig. 8(b), the external magnetic field in Fig. 8(c) remain the same, however, the exciting frequency is changed to $f=0.935f_{\rm p}$. In this condition, the excited EM wave propagates through the waveguide instead of being trapped. It can be understand for the asymptotic frequency $\omega_{\rm sp}^{(1)}=\omega_{\rm c}$ and there are constantly SMPs with $f>\omega_{\rm sp}^{(1)}$ when $\omega_{\rm c}>\omega_0$ (see Fig. 5, Fig. 6(a) and Fig. 7(a)). More interestingly, when enlarge the external magnetic field and let $f/f_{\rm p}=\omega_{\rm c}/\omega_{\rm p}=0.935$, the simulation results in Fig. 8(d) show the excited EM wave is trapped around metal B. In addition, Fig. 8(e) shows the EM waves with higher frequency than $0.935f_{\rm p}$, similar to Fig. 8(c), can not be trapped in the MSSM waveguide when $\omega_{\rm c}=0.935\omega_{\rm p}$. The simulation results shown in Figs. 8(b) and 8(d) fit well with the numerical calculation in Fig. 8(a). 

Due to the ability of trapping EM waves with different frequencies at different location, our designed MSSM waveguide shown in Fig. 8 can be treated as an optical filter, besides, it can also be used as an optical switch or optical buffer. Fig. 9 shows the potential functions of the MSSM structure. First, as an optical filter, the top diagram of Fig. 9 shows that by carefully designing $d_2^{(1)}$ and $d_2^{(2)}$, the EM wave with lower frequency ($f_1$) can be trapped while the EM wave with higher frequency ($f_2$) propagate through the structure. Then, as an optical switch, the last two diagrams of Fig. 9 indicate that the the value of $\omega_{\rm c}$ controls the propagation ('on') or capture ('off') of the EM wave with $f=f_2$. Finally, the function of optical buffer can be achieved based on the function of optical switch by controlling the time difference ($\Delta \tau$) between $\omega_{\rm c}=\omega_{\rm c}^{(1)}$ and $\omega_{\rm c}=\omega_{\rm c}^{(2)}$. The above analysis of optical functional devices can find the according simulations in Fig. 8 when assume $f_1=0.87f_{\rm p}$, $f_2=0.935f_{\rm p}$, $d_2^{(1)}=0.1\lambda_{\rm p}$, $d_2^{(2)}=0.015\lambda_{\rm p}$, $d_2^{(3)}=0.01\lambda_{\rm p}$, $\omega_{\rm c}^{(1)}=0.87\omega_{\rm p}$ and $\omega_{\rm c}^{(2)}=0.935\omega_{\rm p}$. Moreover, due to the one-way propagation properties of the EM waves, our proposed optical functional devices based on the MSSM configuration should possess much higher efficiency than regular ones. 

\section{Conclusion}
In conclusion, we have proposed a novel metal-semiconductor-semiconductor-metal (MSSM) model with only one of the semiconductor layer is under an external magnetic field $B_0$. In our theoretical analysis, we have shown a special three complete one-way propagation (COWP) bands including two COWP bands of surface magnetoplasmons (SMPs) and one COWP band of surface EM modes (SMs) supported by the semiconductor-metal interface. The simultaneously exist one-way SMPs and one-way SMs are promising for designing optical integrated circuit and for the study of nonlocality. Moreover, by controlling the values of $B_0$ and the thicknesses of semiconductors, we found slow-wave valley of the dispersion curves of SMPs in the MSSM waveguide. Accordingly, slow-wave valley-based rainbow trapping have been achieved and, more interestingly, new truly rainbow trapping theory was proposed when the electron cyclotron frequency $\omega_{\rm c}$ being larger than a specific value $\omega_0$ (in this paper, $\omega_0 \approx 0.863\omega_{\rm p}$). By using the finite element method, we further verified the two kinds of rainbow trapping and as a result, the slow-wave valley-based rainbow trapping shows much less reflection than the slow-wave peak-based rainbow trapping reported in our previous work. In the meanwhile, it is the first time that the truly rainbow trapping is achieved in the terahertz regime by simply engineering the thickness of the waveguide. Besides, we also showed the potential uses of our proposed MSSM waveguide in optical filter, optical switch and optical buffer. 

\section*{Funding information}
We acknowledge support by National Natural Science Foundation of China (NSFC) (61865009); National Natural Science Foundation of China (NSFC) (61927813); the Science and Technology Strategic Cooperation Programs of Luzhou Municipal People's Government and Southwest Medical University (2019LZXNYDJ18).


\bibliographystyle{unsrt}
\bibliography{mybib}


\end{document}